\documentclass[aip,jap,numerical,reprint]{revtex4-1}
\usepackage{dcolumn}% Align table columns on decimal point
\usepackage{bm}% bold math
\usepackage{graphicx}
\usepackage{amsmath} 
\usepackage{amssymb}
\usepackage{cleveref}% merge equation reference
\newcommand{\deriv}{\mathrm d}
\begin{document}
\title{Comprehensive Modeling of Graphene Resistivity}
\author{Antonino Contino}
\email{Antonino.Contino@imec.be}
\affiliation{KU Leuven, Electronic department (ESAT), Kasteelpark Arenberg 10, 3001 Leuven, Belgium.}
\affiliation{Imec, Kapeldreef 75, 3001 Leuven, Belgium.}
\author{Ivan~Ciofi}
\affiliation{Imec, Kapeldreef 75, 3001 Leuven, Belgium.}
\author{Xiangyu~Wu}
\affiliation{KU Leuven, Electronic department (ESAT), Kasteelpark Arenberg 10, 3001 Leuven, Belgium.}
\affiliation{Imec, Kapeldreef 75, 3001 Leuven, Belgium.}
\author{Inge~Asselberghs}
\affiliation{Imec, Kapeldreef 75, 3001 Leuven, Belgium.}
\author{Christopher~J.~Wilson}
\affiliation{Imec, Kapeldreef 75, 3001 Leuven, Belgium.}
\author{Zsolt~T\"{o}kei}
\affiliation{Imec, Kapeldreef 75, 3001 Leuven, Belgium.}
\author{Guido~Groeseneken}
\affiliation{KU Leuven, Electronic department (ESAT), Kasteelpark Arenberg 10, 3001 Leuven, Belgium.}
\affiliation{Imec, Kapeldreef 75, 3001 Leuven, Belgium.}
\author{Bart~Sor\'{e}e}
\affiliation{KU Leuven, Electronic department (ESAT), Kasteelpark Arenberg 10, 3001 Leuven, Belgium.}
\affiliation{Imec, Kapeldreef 75, 3001 Leuven, Belgium.}
\affiliation{Universiteit Antwerpen, Physics department, Groenenborgenlaan 171, B-2000 Antwerpen, Belgium.}
\date{\today}
\begin{abstract}
	Since the first graphene layer was fabricated in the early 2000's, graphene properties have been studied extensively both experimentally and theoretically. However, when comparing the many resistivity models reported in literature, several discrepancies can be found, as well as a number of inconsistencies between formulas. In this paper, we revise the main scattering mechanisms in graphene, based on theory and goodness of fit to in-house experimental data. In particular, a step-by-step evaluation of the interaction between electrons and optical phonons is carried out, where we demonstrate that the process of optical phonon emission scattering is completely suppressed for all low-field applications and all temperatures in the range of interest, as opposed to what is often reported in literature. Finally, we identify the best scattering models based on the goodness of fit to experimental data.
\end{abstract}
%
% make the title area
\maketitle
\section{Introduction} \label{Introduction}
Graphene is gaining more and more attention from the scientific community in various academic fields, such as photonics, mechanics, power storage, biomedical sciences and electronics. In the semiconductor industry, graphene is considered as a valid candidate for beyond CMOS not only for transistors \cite{Schwierz2010,Das2008}, but also for interconnects. Indeed, it is now possible to find in literature various studies where graphene is investigated for inteconnect applications as a replacement for the thick Cu barrier or for the Cu itself \cite{Hong2014,Contino2016,Rakheja2013}. In order to predict the performance benefits of graphene interconnects, accurate models calibrated on experimental data are required. 
 
In this paper we provide an overview of several resistivity models, screening the ones available in literature by comparing them with either our own calculations or our measurements. In particular, in Section~\ref{Resistivity Model} the theory of each model is presented, in Section~\ref{Calibration} the models are calibrated to our experimental data, and in Section~\ref{Results} the results of the calibration procedure are discussed. For completeness, Section~\ref{Resistivity Model} also contains a simplified model for the evaluation of edge scattering in graphene, which will be calibrated in future papers. 
\section{Resistivity Model} \label{Resistivity Model}
\subsection{Phonon scattering model} \label{Phonon Scattering}
The intrinsic resistivity of graphene represents the upper limits to its performance. In fact, if in the future the industry is able to produce high quality graphene with no defects, no impurities, perfect edges and to perfectly isolate the graphene from the substrate, phonon scattering will become the only mechanisms able to affect the performance. In literature, although phonon scattering in 2D graphene has been widely investigated, there is still a huge discrepancy on the deformation potentials values for both acoustic and optical phonon scattering \cite{Rakheja2013,Shishir2009,Sule2012}. This discrepancy originates from differences in the formulas used to fit the experimental data, as described in the review paper from Fischetti et al \cite{Fischetti2013}. In the current paper, we also want to bring to the attention that the formulas used to evaluate the scattering rate with optical phonon are often inconsistent. When evaluating optical phonon emission, such formulas from literature \cite{Rakheja2013,Shishir2009,Sule2012,Fischetti2013} assume that the only requirement for an optical phonon to be emitted is that the electron have enough energy, which accordingly is modeled by including in the formula an energy step function. This step function, however, is not capable to describe the complete mechanism on its own. In fact, it is known that the number of electrons in graphene can be enhanced by increasing the voltage applied to a back gate, which finally corresponds to increasing the graphene Fermi energy. Therefore, above a given voltage, the Fermi energy becomes larger than the phonon energy. In this case, the step function would become one and the formula would predict that electrons are able to emit an optical phonon. Several literature plots show indeed a sudden drop in conductivity when either the number of carriers or the Fermi energy are high enough \cite{Contino2016, Rakheja2013,Sule2012,Fischetti2013}. We would like to point out that even if the electrons energy is higher than the phonon energy, there must be available empty states at lower energies for the electrons in order to emit a phonon. This is not the case for the considered example:as the electron energy increases, all the available states below them are simultaneously filled. This effect is captured by the Pauli blocking term, which takes into account the availability of empty states after the scattering event. In the case of quasi-elastic scattering events, such as acoustic phonon scattering, the Pauli blocking term is equal to one and can be neglected. In the case of inelastic scattering events, such as the optical phonon scattering, the Pauli blocking term is not equal to one and must be taken into account when evaluating the scattering rate. In the following, we derive the complete formula for optical phonon scattering and show that the Pauli blocking term completely suppresses the impact of optical phonon emission, even when electrons have higher energy than the phonon energy.

In the Boltzmann transport approach, the graphene conductivity can be evaluated using the relaxation time approximation:
\begin{equation} \label{eq:Boltzmann}
	\sigma=\frac{e^2v_f^2}{2} \int \mkern-3mu \deriv E_k DOS(E_k) \tau (E_k) \left(-\frac{\partial f(E_k)}{\partial E_k} \right) \,,
\end{equation}
where $e$ is the electron charge, $v_f$ is the Fermi velocity, $DOS(E_k)$ is the Density of States in graphene, $f(E_k)$ is the Fermi-Dirac distribution, $E_k$ is the energy as a function of the k vector and $\tau$ is the relaxation time. 

For optical phonon, the relaxation time $\tau_O$ is equal to:
\begin{equation} \label{eq:Fermi}
	\frac{1}{\tau_O} =\sum_\textbf{k'} \Gamma\left(\textbf{k},\textbf{k'}\right)\left(1-\cos \theta\right) B \;,
\end{equation}
\begin{equation} \label{eq:Pauli_term}
	B = \frac{1-f(E(\textbf{k'}))}{1-f(E(\textbf{k}))} \;,
\end{equation}
\begin{equation} \label{eq:Gamma}
	\Gamma\left(\textbf{k},\textbf{k'}\right)=\frac{2\pi}{\hbar} \sum_\textbf{q}| H_{k',k}|^2 \delta(E(\textbf{k'})-E(\textbf{k})\mp \hbar \omega_O) \;,
\end{equation}
where $B$ is the Pauli blocking term, while $\Gamma$ is obtained using Fermi's golden rule, with $\mp$ valid for phonon absorption and emission, respectively. The Hamiltonian can be obtained as
\begin{equation} \label{eq:Hamiltonian}
	H_{k',k} = \int\limits_A \psi_{k'}^*(\textbf{r}) V(\textbf{r}) \psi_{k} d^2r  \;,
\end{equation}
where $\psi_k$ are not simple plane waves, but plane waves spinor solutions of the graphene Dirac Hamiltonian:
\begin{equation} \label{eq:waves}
	\psi_k (\textbf{r}) = \frac{1}{\sqrt{2A}}
	\begin{pmatrix}
	e^{-i\alpha_\textbf{k} / 2 } \\
	e^{i\alpha_\textbf{k} / 2 }
	\end{pmatrix}
	e^{i\textbf{k} \cdot \textbf{r}} \;.
\end{equation}
The potential $V(\textbf{r})$ for absorption and emission has the form
\begin{equation} \label{eq:potential}
	V(\textbf{r})= D_O A_q e^{\pm i(\textbf{q} \cdot \textbf{r} -\omega_O t)} \;,
\end{equation}
with
\begin{equation} \label{eq:Aq}
	|A_q|^2=\frac{\hbar}{2\rho_m A\omega_O} N_\mp \;,
\end{equation}
\begin{equation} \label{eq:number_phonon}
	N_\mp=\frac{1}{e^{\hbar \omega_O / (k_B T)}-1} +\frac{1}{2} \mp \frac{1}{2}= N_q  +\frac{1}{2} \mp \frac{1}{2} \;,
\end{equation}
where $\rho_m$ is the mass density of graphene, $D_O$ is the electron optical deformation potential and $N_q$ is the number of phonons, for which the $\mp$ stands for phonon absorption and emission respectively (the $+1$ is needed to account for the spontaneous emission). By substituting \cref{eq:Aq,eq:number_phonon,eq:potential,eq:waves} in (\ref{eq:Hamiltonian}), solving the integral and evaluating the square modulus, we obtain
\begin{align} \label{eq:Hamiltonian_mod}
	\begin{split}
	|H_{k',k}|^2 &= D_O^2 |A_q|^2 \cos^2(\theta/2) \: \delta_{\textbf{k'},\textbf{k}\pm\textbf{q}} \\ 
	&= D_O^2 \frac{\hbar}{2\rho_m A\omega_O}N_\mp\frac{1+cos(\theta)}{2} \: \delta_{\textbf{k'},\textbf{k}\pm\textbf{q}}
	\end{split}	\;.
\end{align}
Now, we can evaluate the relaxation time using \cref{eq:Fermi,eq:Gamma,eq:Hamiltonian_mod} and the graphene linear dispersion relationship:
\begin{equation}
	\begin{split} 
	\frac{1}{\tau_O} =& \sum_\textbf{k'} \frac{\pi D_O^2}{2\rho_m A \omega_O }N_\mp\sin^2\theta \; \times \\
	& \times\delta(\hbar v_f k' - \hbar v_f k \mp \hbar \omega_O) B
	\end{split}
\end{equation}	
Note that we removed the term $\delta_{\textbf{k'},\textbf{k}\pm\textbf{q}}$ since we assumed $\omega_O$ to be constant, thus the relaxation time does not depend on \textbf{q}. 
	By converting the sum into an integral and normalizing the dirac delta function, we obtain
\begin{align} 
	\frac{1}{\tau_O} =&  \frac{D_O^2N_\mp}{4\pi \rho_m \omega_O } \int\limits_{0}^{2\pi} \!\!  \sin^2\theta d\theta \! \int\limits_0^\infty \!\! \frac{k'}{\hbar v_f } \delta \left(k' - k \mp \frac{\omega_O}{v_f}\right) B dk' \nonumber \\
		=& \frac{D_O^2}{4\rho_m \omega_O \hbar v_f} [ N_q ( k + \frac{\omega_O}{v_f}) B_{Abs} + \label{eq:relax1} \\
		&+ (N_q+1) ( k - \frac{\omega_O}{v_f} )  U(k- \frac{\omega_O}{v_f} ) B_{Em} ] \;, \nonumber
\end{align}
where $U(k- \frac{\omega_O}{v_f} )$ is the step function, while $B_{Em}$ and $B_{Abs}$ are  the Pauli blocking terms for absorption ($E(k')=~E(k)+\hbar \omega_O$) and emission ($E(k')=E(k)-\hbar \omega_O$) respectively. Note that a factor of two has been added in (\ref{eq:relax1}) after considering that, in graphene, the LO and the TO modes give rise to almost the same scattering rate.

Finally, the relaxation time of (\ref{eq:relax1}) can be substituted in (\ref{eq:Boltzmann}) to obtain the exact conductivity for phonon scattering. Alternatively, to obtain a simpler formula, the relaxation time can be approximated by assuming scattering only around the Fermi level ($k=k_f=\sqrt{n \pi}$), thus obtaining the final expression for the relaxation time: 
\begin{align}
	\frac{1}{\tau_O} =& \frac{D_O^2 \sqrt{n \pi}}{4\rho_m \omega_O \hbar v_f} [ N_q ( 1 + \frac{\omega_O}{v_f \sqrt{n \pi}}) B_{Abs-E_f} +  \label{eq:optic_relax}\\
		&+ (N_q+1) ( 1 - \frac{\omega_O}{v_f \sqrt{n \pi}} )  U(k_f- \frac{\omega_O}{v_f} ) B_{Em-E_f} ] \;.\nonumber
\end{align}	
\begin{figure}[!t]
	\centering
	\includegraphics[width=3in]{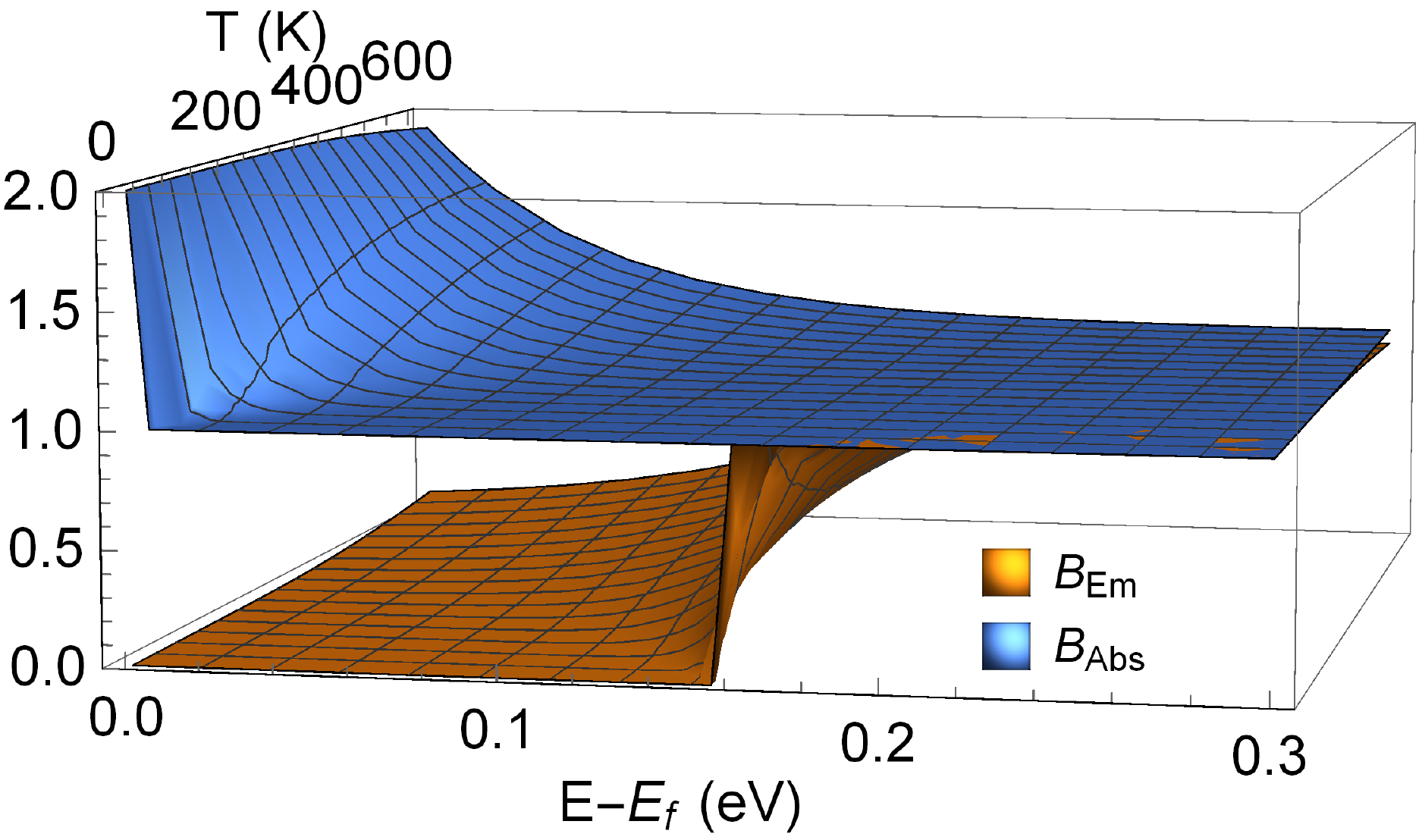}
	\caption{Pauli blocking term for emission (gold) and absorption (blue) as a function of temperature and the difference between the electron energy and the Fermi energy. The phonon energy considered is 160~meV. At energy close to the Fermi energy $(E-E_f=0)$, the factor for emission is $\approx0$ at all temperature, which means that the emission process is completely suppressed.}
	\label{fig:Pauli}
\end{figure}
The equation reported in (\ref{eq:optic_relax}) is equal to the one reported in literature \cite{Fischetti2013,Shishir2009}, except for the Pauli blocking terms. These factors are plotted in Fig.~\ref{fig:Pauli}, where, for completeness, the dependency from both temperature and the difference between the electron energy and the Fermi energy are shown. If (\ref{eq:Boltzmann}) is used, the entire Pauli blocking factor must be considered. If (\ref{eq:optic_relax}) is used, then the factors must be evaluated at the Fermi energy (i.e. $E-E_f=\nolinebreak 0$). Therefore, for absorption, we obtain $B_{Abs-E_f}=\nolinebreak 2$ at $T=\nolinebreak 0$~K, $B_{Abs-E_f}=\nolinebreak 1.996$ at room temperature, and $B_{Abs-E_f}=\nolinebreak 1.913$ at $T=\nolinebreak 600$~K. For emission, we obtain $B_{Em-E_f}=\nolinebreak 0$ at $T=\nolinebreak 0$~K, $B_{Em-E_f}=0.004$ at room temperature, and $B_{Em-E_f}=\nolinebreak 0.087$ at $T=\nolinebreak 600$~K. To summarize, because of the contribution of Pauli blocking, a factor of $\approx 2$ must be considered for the process of optical phonon absorption, while the process of optical phonon emission is completely suppressed for low field applications for all temperatures in the range of interest. Therefore, no conductivity drop is present when increasing either the Fermi energy or the number of carriers. For the sake of completeness, we would like to remark that, for high field applications only, the conductivity drop is still present when plotting the conductivity as a function of the difference between the electron energy and the Fermi energy.

We conclude that the main intrinsic scattering mechanism for graphene is the acoustic phonon scattering, for which the subsequent equation can be derived by following similar calculations, as also shown in literature \cite{Sule2012,Shishir2009,Fischetti2013}. 
\begin{equation} \label{eq:Acoustic_relax}
	\frac{1}{\tau_A} = \frac{D_A^2k_BT \sqrt{n \pi}}{4\rho_m \hbar^2 v^2_s v_f}
\end{equation}
with $v_s$ the sound velocity in graphene and $D_A$ the acoustic deformation potential, which will be calibrated in Section \ref{Calibration}.
\subsection{Long range scattering model} \label{Charged Impurity Scattering}
The main long range scattering mechanisms in graphene originates from the scattering with charged impurities, also known as Coulomb scattering \cite{Chen2008b}. We evaluated it for 2D graphene by means of numerical computations, starting from the theory presented by Adam and Hwang \cite{Hwang2009,Adam2009}. In the following, we summarize those sections of the two papers we referred to for our computations.

The relaxation time associated with long range scattering as a function of the number of impurities $n_{i_L}$ can be evaluated as:
\begin{align} \label{eq:scatt_coul}
	\frac{1}{\tau_L}=\frac{2 \pi n_{i_L}}{\hbar} \int& \mkern-3mu \frac{\deriv^2 k'}{(2\pi)^2} \left| \langle V_{s\textbf{k},s\textbf{k}'}\rangle \right|^2 \times \\
	&\times \left[ 1-\cos \theta_{\textbf{kk}'} \right] \delta \left( E_{s\textbf{k}}-E_{s\textbf{k}'} \right) \nonumber
\end{align} 
To evaluate the scattering rate, we first need to calculate the matrix element $\vert \langle V_{s\textbf{k},s\textbf{k}'}\rangle \vert^2$. If we consider the effect of a point charge at a distance $d$ from the graphene sheet (with $d$ equal to half of the inter-ribbon distance for the case of intercalated charges, or equal to zero for the case of charges in the graphene plane), we can evaluate the matrix element as
\begin{equation}
	\left| \langle V_{s\textbf{k},s\textbf{k}'}\rangle \right|^2 = \left| \frac{V_i(q,d)}{\varepsilon(q,T)} \right|^2 \frac{1+\cos \theta_\textbf{kk'}}{2}
\end{equation}
where $q=\nolinebreak \vert \textbf{k}-\textbf{k}'\vert$, $V_i(q,d)=\nolinebreak 2 \pi e^2 \rm exp(-qd)/(\kappa q)$ is the Fourier transform of the 2D Coulomb potential in an effective background lattice dielectric constant $\kappa$ and $\varepsilon(q,T)$ is the graphene dielectric function, given by $\varepsilon(q,T) =\nolinebreak 1 +\nolinebreak v_c(q)\Pi(q, T)$, where $\Pi (q,T)$ is the graphene polarizability function and $v_c(q)$ is the Coulomb interaction.
The $\Pi (q,T)$ can be evaluated either analytically, in which case assumptions needs to be made, or numerically. By using the calculations of Hwang et al \cite{Hwang2009} up to the point in which assumptions are made, the polarizability can be written as
\begin{equation}
\Pi(q,T)= ( \tilde{\Pi}^+(q,T) + \tilde{\Pi}^-(q,T)) DOS(E_f)
\end{equation}
with
\begin{align} 
	\tilde{\Pi}^+(q,T) = \frac{\mu}{E_f} +& \frac{T}{T_f} \ln\left(1+e^{-\beta \mu} \right) + \label{eq:Piplus} \\
 &- \frac{1}{k_f} \int\limits_0^{q/2} \mkern-3mu \deriv k \frac{\sqrt{1-(2k/q)^2}}{1+e^{\beta(E_k -\mu)}} \nonumber \\
\tilde{\Pi}^-(q,T) = \frac{\pi q}{8 k_f} +& \frac{T}{T_f} \ln\left(1+e^{-\beta \mu} \right) + \label{eq:Piminus} \\ 
&- \frac{1}{k_f} \int\limits_0^{q/2} \mkern-3mu \deriv k \frac{\sqrt{1-(2k/q)^2}}{1+e^{\beta(E_k +\mu)}} \nonumber
\end{align}
In (\ref{eq:Piplus}) and (\ref{eq:Piminus}), $\beta=1/k_bT$ while $\mu$ is the chemical potential, obtained by imposing the charge conservation principle. 
By plugging the previous equations into (\ref{eq:scatt_coul}), the relaxation rate can be rewritten as
\begin{equation}
	\frac{1}{\tau_L}=\frac{n_i E_k}{2\pi k^3 \hbar^3 v_f^2 } \int\limits_0^{2k} \mkern-3mu \deriv q \, q^2 \sqrt{1-\left(\frac{q}{2k}\right)^2}\frac{V_i(q,d)^2}{\varepsilon(q,T)^2}
\end{equation}
This gives the final expression to the relaxation time for Coulomb scattering, to be substituted in (\ref{eq:Boltzmann}).
\subsection{Short range scattering model} \label{Short Range Scattering}
Currently, there are two main theories behind the modeling of scattering from short range potentials such as defects or dislocations in the carbon lattice. The first and more classical one \cite{Hwang2009, Jang2008} follow the exact same calculations of section \ref{Charged Impurity Scattering}, while imposing the graphene polarizability $\Pi (q,T)$ to zero, which corresponds to assume as short range potential a Dirac delta function of amplitude $V_0$. This results in a constant conductivity as a function of the number of carriers at $T=0$~K, equal to:
\begin{equation}	\label{eq:short1}
	\sigma_s = \frac{e^2 vf^2 DOS(E_f) \tau_s}{2\left(1+e^{-\beta \mu}\right)} 
\end{equation}
\begin{equation}	\label{eq:short2}
	 \frac{1}{\tau_s} = \frac{n_{is} E_f V_0^2}{4\hbar^3 v_f^2}
\end{equation} 
Note that (\ref{eq:short2}) is the inverse of what is reported by Hwang et al \cite{Hwang2009}, where we believe a typo error has been made. 
Combining the two equations we obtain:
\begin{equation} \label{eq:short_model1}
 	\sigma_{s}=\frac{4 e^2 v_f^2 \hbar}{\pi n_{is} V_0^2\left(1+e^{-\beta \mu}\right)}
\end{equation} 
However, according to another approach \cite{Peres2010, Stauber2007, Katsnelson2007}, (\ref{eq:short_model1}) is wrong, because when assuming a Dirac delta function for the short range potential in graphene the Born approximation is violated. The right way of describing the short range potential would be, according to this approach, to assume an infinite potential barrier of radius R and calculate the scattering rate using the self-consistent Born approximation. This leads to the following expression for the conductivity:
\begin{equation} \label{eq:short_model2}
	\sigma_s = \frac{2 e^2}{\pi h} \frac{n}{n_{is}} \ln^2(\sqrt{\pi n} R )
\end{equation}
This time, a sub-linear trend as a function of the number of carriers is obtained, even when the radius R is taken as small as the graphene cell size $a_0$.

In the following sections, we apply both (\ref{eq:short_model1}) and (\ref{eq:short_model2}) to understand which of the two approaches fits better the experimental data.
\subsection{Edge scattering model} \label{Edge Scattering}
In the paper of Rakheja et al \cite{Rakheja2013}, a model that allows evaluating the edge scattering resistivity by using the 2D graphene dispersion relationship is reported, considering the quantization of the band structure as a function of the graphene width \textit{W}. Although this is correct for very narrow graphene ribbons, the use of software tools is required, such as Mathematica or Matlab, to obtain the resistivity, since integrals are involved that cannot be solved by hand or have to be evaluated repeatedly for different parameter values; furthermore, the evaluation runtime increases for wider graphene interconnect, because an higher number of sub-bands must be considered. In order to allow an easier and faster evaluation of 2D graphene resistivity down to few tenth of nm, we implemented a simpler model, from which we derived a simple formula, that can be easily implemented in an excel worksheet. 
\begin{figure}[!t]
	\centering
	\includegraphics[width=3in]{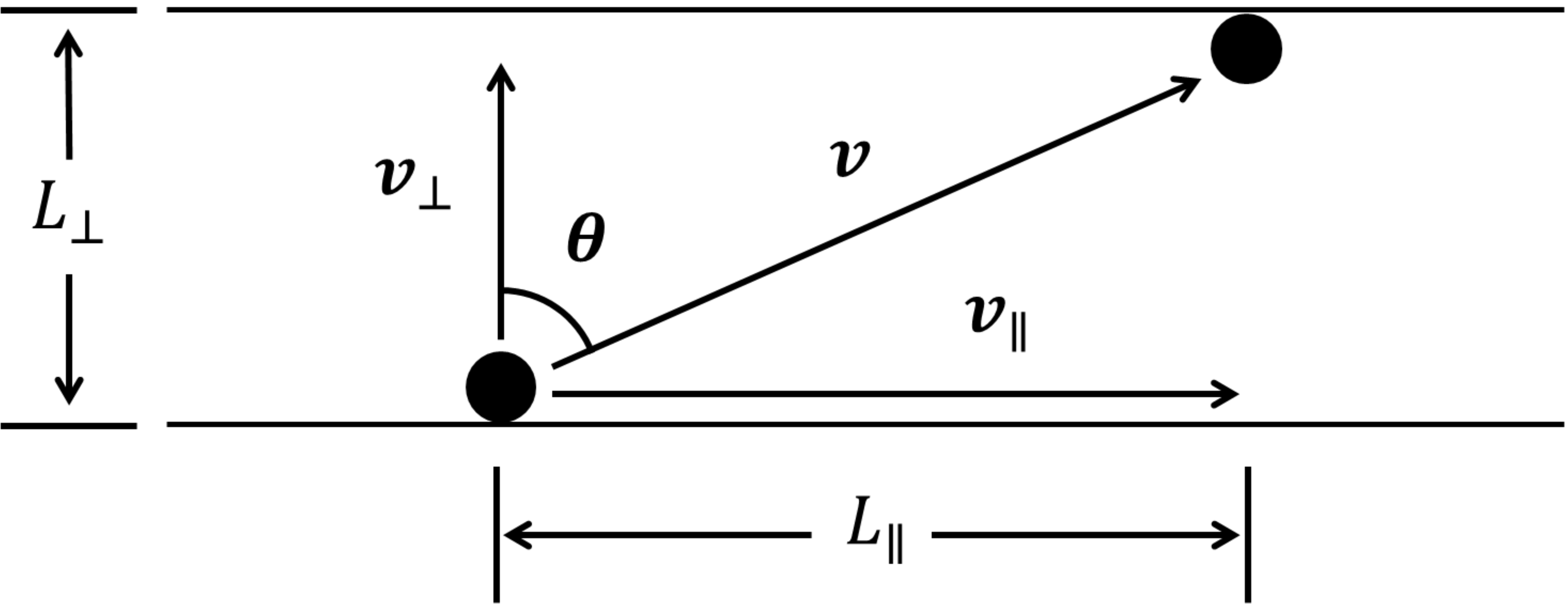}
	\caption{Simplified scheme describing our model for the evaluation of edge scattering in graphene. After impacting with one edge, the electron will cover a distance $L_\parallel$ along the transport direction before impacting again with the other edge.}
	\label{fig:model1}
\end{figure}
Let's consider an electron traveling with velocity \textbf{v} in the graphene, immediately after impacting with one of the graphene edges (see Fig.~\ref{fig:model1}). Before impacting again with the other edge, it will cover a distance $L_\parallel$ along the transport direction and a distance $L_\perp$ along the direction perpendicular to transport, that can be evaluated as:
\begin{equation} 
	\bigg \{ \begin{matrix}
	L_\parallel = v_\parallel \tau = |\textbf{v}| \sin(\theta) \tau \\
	L_\perp = v_\perp \tau = |\textbf{v}| \cos(\theta) \tau
	\end{matrix} 
\end{equation}
Where $\tau$ is the time and $\theta$ is the angle of the vector \textbf{v}. By simply solving the system with respect to $L_\parallel$  we obtain
\begin{equation} \label{eq:Lparal}
	L_\parallel =  L_\perp \frac{|\textbf{v}|\sin(\theta)}{|\textbf{v}|\cos(\theta)}  = W \tan(\theta) 
\end{equation}
in which we considered that $L_\perp$ is approximately equal to the graphene width W. After the impact, depending on the amount of edge roughness, there is a probability P of diffusive scattering, which reduces the Mean Free Path (MFP) in the transport direction, and a probability $(1-P)$ of specular scattering, which does not affect the performance. Therefore, we can evaluate the MFP associated with edge scattering by averaging (\ref{eq:Lparal}) with respect to $\theta$, including the diffusive scattering probability and normalizing the equation.
\begin{equation} \label{eq:lambdaedge}
	\lambda_\textrm{edge}=\frac{2}{\pi} \int\limits_0^{\pi/2} \frac{1}{P} L_\parallel \deriv\theta=\frac{2}{\pi} \int\limits_0^{\pi/2} \frac{1}{P} W \tan(\theta) \deriv\theta
\end{equation}
Note that we integrated only between 0 and $\frac{\pi}{2}$ thanks to the symmetry of the system. Unfortunately, (\ref{eq:lambdaedge}) does not converge: for $\theta=\frac{\pi}{2}$, the electron will never impact with the edges again, thus an infinite MFP is obtained, independently from the other parameters. However, in reality, the electron will not travel indefinitely with $\theta=\frac{\pi}{2}$; other scattering mechanisms will come into play, reducing the MFP and changing the electron angle again (e.g. phonon scattering, short range scattering, etc.). Therefore, we can re-write (\ref{eq:lambdaedge}) in therms of total MFP as:
\begin{align} \label{eq:edge_lambdatot}
	\lambda_{tot} &=\frac{2}{\pi} \int\limits_0^{\pi/2} \frac{1}{\frac{P}{L_\parallel}+\frac{1}{\lambda_{Max}}} \deriv\theta = \frac{2}{\pi} \int\limits_0^{\pi/2} \frac{1}{\frac{P}{ W tan(\theta)}+\frac{1}{\lambda_{Max}}} \deriv\theta= \nonumber \\ 
	&=  \frac{W \lambda_{Max} \left( \pi W + 2 P \lambda_{Max} \ln[\frac{P \lambda_{Max}}{W}] \right)}{\pi \left( W^2+P^2  \lambda_{Max}^2 \right)}
\end{align}
This way, for $\theta=\frac{\pi}{2}$, $P=0$ or for very wide graphene, the total MFP is simply equal to the MFP associated with the other scattering mechanisms, $\lambda_{Max}$. Finally, we can evaluate the mobility and the conductivity using the Boltzmann theory for diffusive transport and the graphene density of states.
\begin{equation} 
	\mu_{tot}= \frac{eDOS(E_f)v_f}{2n}\lambda_{tot}=\frac{e}{\sqrt{n\pi}\hbar}\lambda_{tot}
\end{equation}
\begin{equation}
	\sigma_{tot} = e n \mu_{tot}.
\end{equation}
\section{Calibration} \label{Calibration}
\subsection{Calibration method}	\label{Method}
We calibrated our resistivity model to the experimental data presented by Wu et al \cite{Wu2017a}, which consist of in-house multi-temperature measurements of a CVD grown graphene ribbon of $5~\mu$m width, performed in vacuum. Since the characterized ribbon is very large, it is reasonable to assume the impact of edge scattering to be negligible. Therefore, the measured conductivity data from \cite{Wu2017a} are not suitable for calibrating our edge scattering model, which will be calibrated in another paper.

For the other scattering models, the calibration procedure is divided in three steps: firstly, we calibrate the models for long and short range scattering using the measurements at low temperature, where phonon contribution can be neglected; secondly, we calibrate the phonon model using another set of measurements at higher temperature; thirdly, we use the remaining measurements to validate the temperature trend of our models. This validation is possible only if the number of impurities for both short and long range scattering remains unchanged with temperature. Unfortunately, this is not the case for the measurements at $T=300$~K \cite{Wu2017a}, for which we believe that water intercalation, frozen at lower temperature but melted at room temperature, causes an increase in both doping level (i.e. shift of the neutrality point) and impurity level. Thus, this set of measurements is not used in this calibration. 

To summarize, we use the measurements at $T=9.6$~K to calibrate the long and short range scattering models, the measurements at $T=200$~K to calibrate the phonon scattering model, and finally the measurements at $T=100$~K to validate the temperature trend predicted by the calibrated models. The three measurements sets are shown in Fig.~\ref{fig:Final_Calib}.
\subsection{Long and short range scattering calibration} \label{Long and Short Range Scattering}
The calibration procedure for long and short range scattering differ depending on the model used for short range scattering. 

If (\ref{eq:short_model1}) is used, the short range scattering conductivity is constant with the number of carrier. Thus, it is easy to separate the contribution of long range (linear conductivity with the number of carriers) and short range. In fact, using Matthiessen's rule we have that:
\begin{equation} \label{eq:calibr_short_long}
\rho_{tot} (n) = \frac{1}{\sigma_{long} (n)} + \frac{1}{\sigma_{short}} \propto \frac{1}{n} + const
\end{equation}
Therefore, we can first find which constant value we need to remove from the measured resistivity in order to obtain a $\frac{1}{n}$ dependency. This constant value will be used to calibrate the short range model. Next, we can calibrate the number of long range impurities by fitting the model to the measured resistivity without the contribution of short range. 

Using the measurements at $T=9.6$~K, we extracted a constant conductivity $\sigma_{short}=16.27$~mS, from which we obtain, using (\ref{eq:short_model1}), a number of impurity for short range scattering $n_{is}=7 \times 10^{12} \rm cm^{-2}$. Then, by fitting the measurements using (\ref{eq:scatt_coul}) and (\ref{eq:calibr_short_long}), we can extract the number of impurity for coulomb scattering $n_{ic}=1.44 \times 10^{12} \rm cm^{-2}$.

If (\ref{eq:short_model2}) is used instead, then the two contribution cannot be separated anymore. By fitting the number of impurities for both models at the same time using the least square method, we obtain $n_{is}=4.65 \times 10^{10} \rm cm^{-2}$ and $n_{ic}=1.07 \times 10^{12} \rm cm^{-2}$. Note that with this model the number of impurity extracted for short range scattering is orders of magnitude lower than in the previous case.
\subsection{Phonon scattering calibration}
If we compare the measurement at $T=200$~K with the one at $T=9.6$~K, we know from theory that the change in conductivity is related to the contribution of phonon scattering (mostly, because also the long range scattering has a small temperature dependency). If we write the equation for the resistivity using Mattiessen's rule, we obtain
\begin{align}
	\rho_{200K} &= \frac{1}{ne} \left(\frac{1}{\mu_{ph}} + \frac{1}{\mu_{l_{200K}}} + \frac{1}{\mu_s} \right) \label{eq:phon_200_res}\\
	\rho_{9.6K} &= \frac{1}{ne} \left( \frac{1}{\mu_{l_{9.6K}}} + \frac{1}{\mu_s} \right) \label{eq:phon_10_res}
\end{align}
where $\mu_{ph}$, $\mu_{l}$ and $\mu_{s}$ are the mobility associated with phonon scattering, long range scattering and short range scattering respectively. If we now calculate the difference between (\ref{eq:phon_200_res}) and (\ref{eq:phon_10_res}) and we solve for $\mu_{ph}$ we obtain:
\begin{equation} \label{eq:calibr_phonon}
	\mu_{ph}=\frac{1}{ne(\rho_{200K}-\rho_{9.6K})+\frac{\mu_{l_{200K}}-\mu_{l_{9.6K}}}{\mu_{l_{200K}} \mu_{l_{9.6K}}}}
\end{equation}
\begin{figure}[!t]
	\centering
	\includegraphics[width=3in]{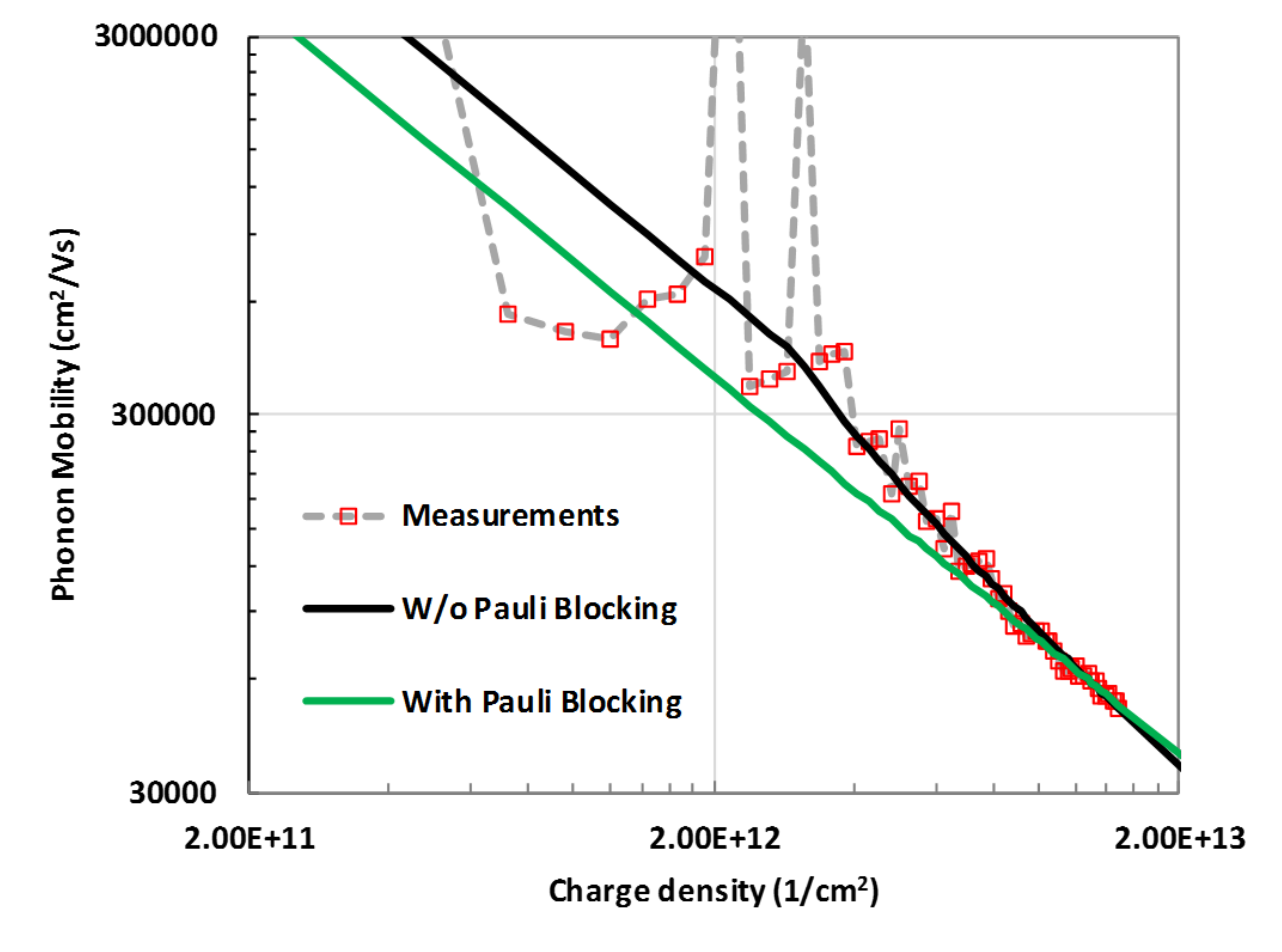}
	\caption{Mobility associated with phonon scattering. The black and green lines represents the literature model and our model, respectively; the squares are the values extracted from the measurements using (\ref{eq:calibr_phonon}), while the grey dashed line linearly connects the measurements and it has been added only to show the spikes in the measured mobility values at low charge density.}
	\label{fig:Calib_phonon}
\end{figure}
Fig.~\ref{fig:Calib_phonon} shows the mobility obtained by using (\ref{eq:calibr_phonon}), where the red squares represents the value extracted from the measurements, the grey dashed line linearly connects the measurements, the black line is obtained by using the phonon models from literature and the green line is the one we obtained using our models with the Pauli blocking term for the optical phonon. Note that the mobility extracted from the measurements shows huge variation at low charge density, due to the error in the estimation of the exact number of charges near the neutrality point. This error, combined with the low impact of phonon scattering at low charge density, results in a small difference between the two measurements, which causes spikes in the extracted mobility. However, for high charge densities, the measurements are more accurate and thus the models must be fitted using mostly this part.
The deformation potentials for acoustic and optical phonon extracted by using the literature models are $D_A = 6.5$~eV and  $D_O = 3 \times 10^{10} \rm eVm^{-1}$ respectively. By using our model, the impact of optical phonon absorption is so small that we were only able to fit the deformation potential for acoustic phonon, $D_A=8.4$~eV. 
\begin{figure}[!t]
	\centering
	\includegraphics[width=3.37in]{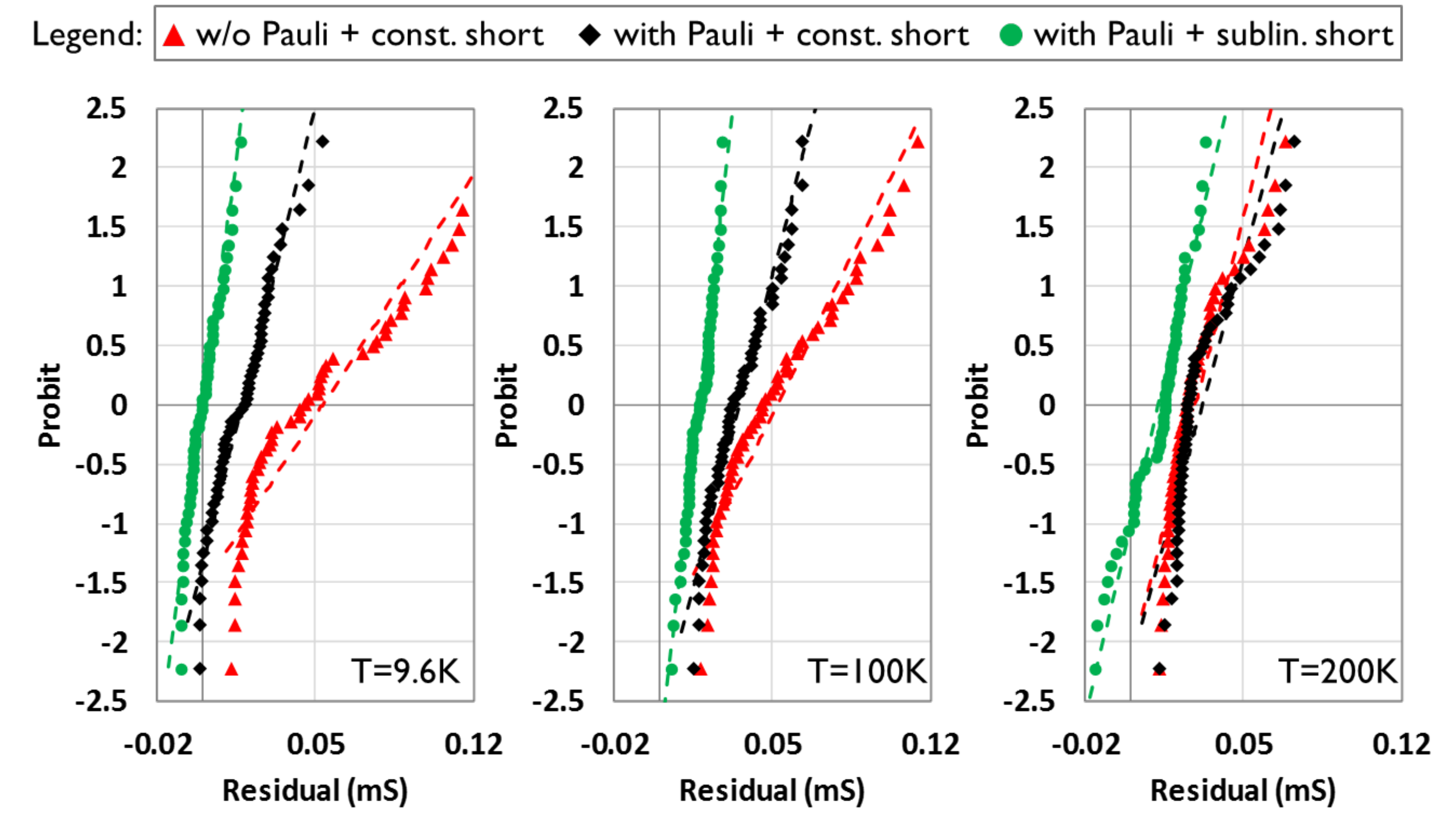}
	\caption{Residuals plots at three different temperature for the three cases considered. The green distribution, which uses (\ref{eq:short_model2}) for short range scattering (sublinear) and (\ref{eq:optic_relax}) and (\ref{eq:Acoustic_relax}) for phonon scattering (including Pauli blocking), fits better the experimental data: it has the lowest mean error (intercept), the lowest variability (slope) and a negligible deviation from the linear trend.}
	\label{fig:Residuals}
\end{figure}
\begin{figure}[!t]
	\centering
	\includegraphics[width=3.37in]{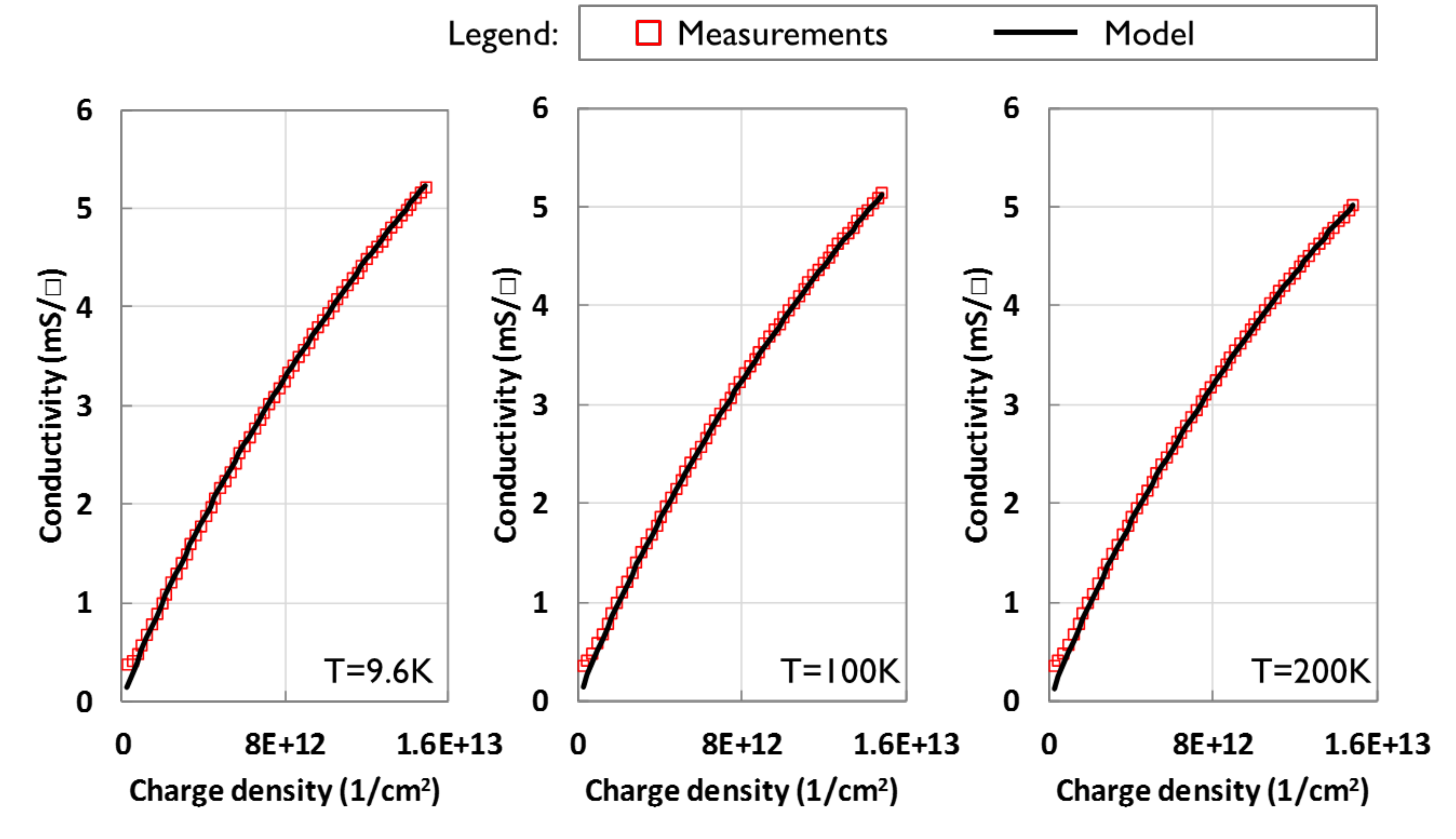}
	\caption{Comparison between the measured conductivity (red squares) and the prediction made using our champion model (black solid line) for the three different temperature considered. A good match between model-based predictions and experimental data is obtained.}
	\label{fig:Final_Calib}
\end{figure}
\section{Results} \label{Results}
For each of the investigated models, we evaluated the residuals ($\Delta=\sigma_{Meas}-\sigma_{Model}$) at each temperature and plotted them on a probit scale (Fig.~\ref{fig:Residuals}). In Fig.~\ref{fig:Residuals}, the investigated models can be compared according to three criteria: mean error, given by the intercept between the curves and the x-axis; variability, obtained considering the slope of the curve; deviation from the Gaussian distribution, given by the deviation from the linear trend. The red triangles are obtained by considering the literature phonon scattering model (without Pauli blocking) and the constant behavior for short range scattering, the black diamonds are obtained considering our phonon scattering model (with Pauli blocking) and constant behavior for short range scattering, and finally the green dots are obtained using our phonon scattering model and the sub-linear behavior for short range scattering. It is clear from Fig.~\ref{fig:Residuals} that the red distribution (i.e. w/o Pauli blocking, constant short) is the worst of the three cases: it has the highest mean error, which increases at low temperature due to the phonon contribution, the highest variability and it has huge deviations from the linear trend. The black distribution (i.e. with Pauli blocking, constant short) gives already a better fit to the experiment. However, the best fit is achieved by the green distribution (i.e. with Pauli blocking, sub-linear short), which show the lowest mean error, the lowest variability and a negligible deviation from the linear trend.  For completeness, Fig.~\ref{fig:Final_Calib} shows both the prediction based on this last model and the considered measurements, where it can be seen that model-based predictions match measurements data, even for the set of measurements that was not used in the calibration procedure at $T=100$~K, which can be regarded as a validation of our modeling and calibration.  
\section*{Conclusion}
Graphene resistivity models were revised based on theory and goodness of fit to experimental data. A more accurate formula for the evaluation of optical phonon scattering was derived and it was proven that the scattering from optical phonon is completely suppressed for all low field applications and for temperature up to 600~K and beyond. The validity of our formula is further confirmed by the significantly better fit to the experimental data we were able to achieve. Moreover, at low temperature, where phonon do not contribute, our formula remains consistent, as opposed to models from literature. Finally, two different literature models for short range scattering in graphene were compared. It was found that a sublinear description of this phenomena \cite{Peres2010, Stauber2007, Katsnelson2007} gives a much better fit of the experimental data. As an outlook, a new calibration procedure must be implemented for our simplified edge scattering model, in order to obtain reliable prediction of graphene resistivity at scaled dimensions.
%
%
%\begin{acknowledgments}
% The author would like to thank...
%\end{acknowledgments}
%
%
\bibliography{Comprehensive_Modeling}

%merlin.mbs aipnum4-1.bst 2010-07-25 4.21a (PWD, AO, DPC) hacked
%Control: key (0)
%Control: author (8) initials jnrlst
%Control: editor formatted (1) identically to author
%Control: production of article title (-1) disabled
%Control: page (0) single
%Control: year (1) truncated
%Control: production of eprint (0) enabled
\begin{thebibliography}{16}%
\makeatletter
\providecommand \@ifxundefined [1]{%
 \@ifx{#1\undefined}
}%
\providecommand \@ifnum [1]{%
 \ifnum #1\expandafter \@firstoftwo
 \else \expandafter \@secondoftwo
 \fi
}%
\providecommand \@ifx [1]{%
 \ifx #1\expandafter \@firstoftwo
 \else \expandafter \@secondoftwo
 \fi
}%
\providecommand \natexlab [1]{#1}%
\providecommand \enquote  [1]{``#1''}%
\providecommand \bibnamefont  [1]{#1}%
\providecommand \bibfnamefont [1]{#1}%
\providecommand \citenamefont [1]{#1}%
\providecommand \href@noop [0]{\@secondoftwo}%
\providecommand \href [0]{\begingroup \@sanitize@url \@href}%
\providecommand \@href[1]{\@@startlink{#1}\@@href}%
\providecommand \@@href[1]{\endgroup#1\@@endlink}%
\providecommand \@sanitize@url [0]{\catcode `\\12\catcode `\$12\catcode
  `\&12\catcode `\#12\catcode `\^12\catcode `\_12\catcode `\%12\relax}%
\providecommand \@@startlink[1]{}%
\providecommand \@@endlink[0]{}%
\providecommand \url  [0]{\begingroup\@sanitize@url \@url }%
\providecommand \@url [1]{\endgroup\@href {#1}{\urlprefix }}%
\providecommand \urlprefix  [0]{URL }%
\providecommand \Eprint [0]{\href }%
\providecommand \doibase [0]{http://dx.doi.org/}%
\providecommand \selectlanguage [0]{\@gobble}%
\providecommand \bibinfo  [0]{\@secondoftwo}%
\providecommand \bibfield  [0]{\@secondoftwo}%
\providecommand \translation [1]{[#1]}%
\providecommand \BibitemOpen [0]{}%
\providecommand \bibitemStop [0]{}%
\providecommand \bibitemNoStop [0]{.\EOS\space}%
\providecommand \EOS [0]{\spacefactor3000\relax}%
\providecommand \BibitemShut  [1]{\csname bibitem#1\endcsname}%
\let\auto@bib@innerbib\@empty
%</preamble>
\bibitem [{\citenamefont {Schwierz}(2010)}]{Schwierz2010}%
  \BibitemOpen
  \bibfield  {author} {\bibinfo {author} {\bibfnamefont {F.}~\bibnamefont
  {Schwierz}},\ }\href {\doibase 10.1038/nnano.2010.89} {\bibfield  {journal}
  {\bibinfo  {journal} {Nature Nanotechnology}\ }\textbf {\bibinfo {volume}
  {5}},\ \bibinfo {pages} {487} (\bibinfo {year} {2010})},\ \Eprint
  {http://arxiv.org/abs/NIHMS150003} {arXiv:NIHMS150003} \BibitemShut {NoStop}%
\bibitem [{\citenamefont {Das}\ \emph {et~al.}(2008)\citenamefont {Das},
  \citenamefont {Pisana}, \citenamefont {Chakraborty}, \citenamefont
  {Piscanec}, \citenamefont {Saha}, \citenamefont {Waghmare}, \citenamefont
  {Novoselov}, \citenamefont {Krishnamurthy}, \citenamefont {Geim},
  \citenamefont {Ferrari},\ and\ \citenamefont {Sood}}]{Das2008}%
  \BibitemOpen
  \bibfield  {author} {\bibinfo {author} {\bibfnamefont {a.}~\bibnamefont
  {Das}}, \bibinfo {author} {\bibfnamefont {S.}~\bibnamefont {Pisana}},
  \bibinfo {author} {\bibfnamefont {B.}~\bibnamefont {Chakraborty}}, \bibinfo
  {author} {\bibfnamefont {S.}~\bibnamefont {Piscanec}}, \bibinfo {author}
  {\bibfnamefont {S.~K.}\ \bibnamefont {Saha}}, \bibinfo {author}
  {\bibfnamefont {U.~V.}\ \bibnamefont {Waghmare}}, \bibinfo {author}
  {\bibfnamefont {K.~S.}\ \bibnamefont {Novoselov}}, \bibinfo {author}
  {\bibfnamefont {H.~R.}\ \bibnamefont {Krishnamurthy}}, \bibinfo {author}
  {\bibfnamefont {a.~K.}\ \bibnamefont {Geim}}, \bibinfo {author}
  {\bibfnamefont {a.~C.}\ \bibnamefont {Ferrari}}, \ and\ \bibinfo {author}
  {\bibfnamefont {a.~K.}\ \bibnamefont {Sood}},\ }\href {\doibase
  10.1038/nnano.2008.67} {\bibfield  {journal} {\bibinfo  {journal} {Nature
  nanotechnology}\ }\textbf {\bibinfo {volume} {3}},\ \bibinfo {pages} {210}
  (\bibinfo {year} {2008})},\ \Eprint {http://arxiv.org/abs/0709.1174}
  {arXiv:0709.1174} \BibitemShut {NoStop}%
\bibitem [{\citenamefont {Hong}\ \emph {et~al.}(2014)\citenamefont {Hong},
  \citenamefont {Lee}, \citenamefont {Lee}, \citenamefont {Han}, \citenamefont
  {Mahata}, \citenamefont {Yeon}, \citenamefont {Koo}, \citenamefont {Kim},
  \citenamefont {Nam}, \citenamefont {Byun}, \citenamefont {Min}, \citenamefont
  {Kim}, \citenamefont {Kim}, \citenamefont {Joo},\ and\ \citenamefont
  {Lee}}]{Hong2014}%
  \BibitemOpen
  \bibfield  {author} {\bibinfo {author} {\bibfnamefont {J.}~\bibnamefont
  {Hong}}, \bibinfo {author} {\bibfnamefont {S.}~\bibnamefont {Lee}}, \bibinfo
  {author} {\bibfnamefont {S.}~\bibnamefont {Lee}}, \bibinfo {author}
  {\bibfnamefont {H.}~\bibnamefont {Han}}, \bibinfo {author} {\bibfnamefont
  {C.}~\bibnamefont {Mahata}}, \bibinfo {author} {\bibfnamefont {H.-W.}\
  \bibnamefont {Yeon}}, \bibinfo {author} {\bibfnamefont {B.}~\bibnamefont
  {Koo}}, \bibinfo {author} {\bibfnamefont {S.-I.}\ \bibnamefont {Kim}},
  \bibinfo {author} {\bibfnamefont {T.}~\bibnamefont {Nam}}, \bibinfo {author}
  {\bibfnamefont {K.}~\bibnamefont {Byun}}, \bibinfo {author} {\bibfnamefont
  {B.-W.}\ \bibnamefont {Min}}, \bibinfo {author} {\bibfnamefont {Y.-W.}\
  \bibnamefont {Kim}}, \bibinfo {author} {\bibfnamefont {H.}~\bibnamefont
  {Kim}}, \bibinfo {author} {\bibfnamefont {Y.-C.}\ \bibnamefont {Joo}}, \ and\
  \bibinfo {author} {\bibfnamefont {T.}~\bibnamefont {Lee}},\ }\href {\doibase
  10.1039/C3NR06771H} {\bibfield  {journal} {\bibinfo  {journal} {Nanoscale}\
  }\textbf {\bibinfo {volume} {6}},\ \bibinfo {pages} {7503} (\bibinfo {year}
  {2014})}\BibitemShut {NoStop}%
\bibitem [{\citenamefont {Contino}\ \emph {et~al.}(2016)\citenamefont
  {Contino}, \citenamefont {Ciofi}, \citenamefont {Politou}, \citenamefont
  {Verkest}, \citenamefont {Mocuta}, \citenamefont {Sor{\'{e}}e},\ and\
  \citenamefont {Groeseneken}}]{Contino2016}%
  \BibitemOpen
  \bibfield  {author} {\bibinfo {author} {\bibfnamefont {A.}~\bibnamefont
  {Contino}}, \bibinfo {author} {\bibfnamefont {I.}~\bibnamefont {Ciofi}},
  \bibinfo {author} {\bibfnamefont {M.}~\bibnamefont {Politou}}, \bibinfo
  {author} {\bibfnamefont {D.}~\bibnamefont {Verkest}}, \bibinfo {author}
  {\bibfnamefont {D.}~\bibnamefont {Mocuta}}, \bibinfo {author} {\bibfnamefont
  {B.}~\bibnamefont {Sor{\'{e}}e}}, \ and\ \bibinfo {author} {\bibfnamefont
  {G.}~\bibnamefont {Groeseneken}},\ }\href {\doibase
  10.1109/IITC-AMC.2016.7507675} {\bibfield  {journal} {\bibinfo  {journal}
  {2016 IEEE International Interconnect Technology Conference / Advanced
  Metallization Conference, IITC/AMC 2016}\ ,\ \bibinfo {pages} {45}} (\bibinfo
  {year} {2016})}\BibitemShut {NoStop}%
\bibitem [{\citenamefont {Rakheja}, \citenamefont {Kumar},\ and\ \citenamefont
  {Naeemi}(2013)}]{Rakheja2013}%
  \BibitemOpen
  \bibfield  {author} {\bibinfo {author} {\bibfnamefont {S.}~\bibnamefont
  {Rakheja}}, \bibinfo {author} {\bibfnamefont {V.}~\bibnamefont {Kumar}}, \
  and\ \bibinfo {author} {\bibfnamefont {A.}~\bibnamefont {Naeemi}},\ }\href
  {\doibase 10.1109/JPROC.2013.2260235} {\bibfield  {journal} {\bibinfo
  {journal} {Proceedings of the IEEE}\ }\textbf {\bibinfo {volume} {101}},\
  \bibinfo {pages} {1740} (\bibinfo {year} {2013})}\BibitemShut {NoStop}%
\bibitem [{\citenamefont {Shishir}\ and\ \citenamefont
  {Ferry}(2009)}]{Shishir2009}%
  \BibitemOpen
  \bibfield  {author} {\bibinfo {author} {\bibfnamefont {R.~S.}\ \bibnamefont
  {Shishir}}\ and\ \bibinfo {author} {\bibfnamefont {D.~K.}\ \bibnamefont
  {Ferry}},\ }\href {\doibase 10.1088/0953-8984/21/23/232204} {\bibfield
  {journal} {\bibinfo  {journal} {Journal of physics. Condensed matter : an
  Institute of Physics journal}\ }\textbf {\bibinfo {volume} {21}},\ \bibinfo
  {pages} {232204} (\bibinfo {year} {2009})}\BibitemShut {NoStop}%
\bibitem [{\citenamefont {Sule}\ and\ \citenamefont
  {Knezevic}(2012)}]{Sule2012}%
  \BibitemOpen
  \bibfield  {author} {\bibinfo {author} {\bibfnamefont {N.}~\bibnamefont
  {Sule}}\ and\ \bibinfo {author} {\bibfnamefont {I.}~\bibnamefont
  {Knezevic}},\ }\href {\doibase 10.1063/1.4747930} {\bibfield  {journal}
  {\bibinfo  {journal} {Journal of Applied Physics}\ }\textbf {\bibinfo
  {volume} {112}} (\bibinfo {year} {2012}),\ 10.1063/1.4747930}\BibitemShut
  {NoStop}%
\bibitem [{\citenamefont {Fischetti}\ \emph {et~al.}(2013)\citenamefont
  {Fischetti}, \citenamefont {Kim}, \citenamefont {Narayanan}, \citenamefont
  {Ong}, \citenamefont {Sachs}, \citenamefont {Ferry},\ and\ \citenamefont
  {Aboud}}]{Fischetti2013}%
  \BibitemOpen
  \bibfield  {author} {\bibinfo {author} {\bibfnamefont {M.~V.}\ \bibnamefont
  {Fischetti}}, \bibinfo {author} {\bibfnamefont {J.}~\bibnamefont {Kim}},
  \bibinfo {author} {\bibfnamefont {S.}~\bibnamefont {Narayanan}}, \bibinfo
  {author} {\bibfnamefont {Z.-Y.}\ \bibnamefont {Ong}}, \bibinfo {author}
  {\bibfnamefont {C.}~\bibnamefont {Sachs}}, \bibinfo {author} {\bibfnamefont
  {D.~K.}\ \bibnamefont {Ferry}}, \ and\ \bibinfo {author} {\bibfnamefont
  {S.~J.}\ \bibnamefont {Aboud}},\ }\href {\doibase
  10.1088/0953-8984/25/47/473202} {\bibfield  {journal} {\bibinfo  {journal}
  {Journal of physics. Condensed matter : an Institute of Physics journal}\
  }\textbf {\bibinfo {volume} {25}},\ \bibinfo {pages} {473202} (\bibinfo
  {year} {2013})}\BibitemShut {NoStop}%
\bibitem [{\citenamefont {Chen}\ \emph {et~al.}(2008)\citenamefont {Chen},
  \citenamefont {Jang}, \citenamefont {Adam}, \citenamefont {Fuhrer},
  \citenamefont {Williams},\ and\ \citenamefont {Ishigami}}]{Chen2008b}%
  \BibitemOpen
  \bibfield  {author} {\bibinfo {author} {\bibfnamefont {J.~H.}\ \bibnamefont
  {Chen}}, \bibinfo {author} {\bibfnamefont {C.}~\bibnamefont {Jang}}, \bibinfo
  {author} {\bibfnamefont {S.}~\bibnamefont {Adam}}, \bibinfo {author}
  {\bibfnamefont {M.~S.}\ \bibnamefont {Fuhrer}}, \bibinfo {author}
  {\bibfnamefont {E.~D.}\ \bibnamefont {Williams}}, \ and\ \bibinfo {author}
  {\bibfnamefont {M.}~\bibnamefont {Ishigami}},\ }\href {\doibase
  10.1038/nphys935} {\bibfield  {journal} {\bibinfo  {journal} {Nature
  Physics}\ }\textbf {\bibinfo {volume} {4}},\ \bibinfo {pages} {377} (\bibinfo
  {year} {2008})},\ \Eprint {http://arxiv.org/abs/0708.2408} {arXiv:0708.2408
  [cond-mat.other]} \BibitemShut {NoStop}%
\bibitem [{\citenamefont {Hwang}\ and\ \citenamefont {{Das
  Sarma}}(2009)}]{Hwang2009}%
  \BibitemOpen
  \bibfield  {author} {\bibinfo {author} {\bibfnamefont {E.~H.}\ \bibnamefont
  {Hwang}}\ and\ \bibinfo {author} {\bibfnamefont {S.}~\bibnamefont {{Das
  Sarma}}},\ }\href {\doibase 10.1103/PhysRevB.79.165404} {\bibfield  {journal}
  {\bibinfo  {journal} {Physical Review B - Condensed Matter and Materials
  Physics}\ }\textbf {\bibinfo {volume} {79}},\ \bibinfo {pages} {1} (\bibinfo
  {year} {2009})},\ \Eprint {http://arxiv.org/abs/0811.1212} {arXiv:0811.1212}
  \BibitemShut {NoStop}%
\bibitem [{\citenamefont {Adam}\ \emph {et~al.}(2009)\citenamefont {Adam},
  \citenamefont {Hwang}, \citenamefont {Rossi},\ and\ \citenamefont {{Das
  Sarma}}}]{Adam2009}%
  \BibitemOpen
  \bibfield  {author} {\bibinfo {author} {\bibfnamefont {S.}~\bibnamefont
  {Adam}}, \bibinfo {author} {\bibfnamefont {E.~H.}\ \bibnamefont {Hwang}},
  \bibinfo {author} {\bibfnamefont {E.}~\bibnamefont {Rossi}}, \ and\ \bibinfo
  {author} {\bibfnamefont {S.}~\bibnamefont {{Das Sarma}}},\ }\href {\doibase
  10.1016/j.ssc.2009.02.041} {\bibfield  {journal} {\bibinfo  {journal} {Solid
  State Communications}\ }\textbf {\bibinfo {volume} {149}},\ \bibinfo {pages}
  {1072} (\bibinfo {year} {2009})},\ \Eprint {http://arxiv.org/abs/0812.1795}
  {arXiv:0812.1795} \BibitemShut {NoStop}%
\bibitem [{\citenamefont {Jang}\ \emph {et~al.}(2008)\citenamefont {Jang},
  \citenamefont {Adam}, \citenamefont {Chen}, \citenamefont {Williams},
  \citenamefont {{Das Sarma}},\ and\ \citenamefont {Fuhrer}}]{Jang2008}%
  \BibitemOpen
  \bibfield  {author} {\bibinfo {author} {\bibfnamefont {C.}~\bibnamefont
  {Jang}}, \bibinfo {author} {\bibfnamefont {S.}~\bibnamefont {Adam}}, \bibinfo
  {author} {\bibfnamefont {J.~H.}\ \bibnamefont {Chen}}, \bibinfo {author}
  {\bibfnamefont {E.~D.}\ \bibnamefont {Williams}}, \bibinfo {author}
  {\bibfnamefont {S.}~\bibnamefont {{Das Sarma}}}, \ and\ \bibinfo {author}
  {\bibfnamefont {M.~S.}\ \bibnamefont {Fuhrer}},\ }\href {\doibase
  10.1103/PhysRevLett.101.146805} {\bibfield  {journal} {\bibinfo  {journal}
  {Physical Review Letters}\ }\textbf {\bibinfo {volume} {101}},\ \bibinfo
  {pages} {1} (\bibinfo {year} {2008})},\ \Eprint
  {http://arxiv.org/abs/0805.3780} {arXiv:0805.3780} \BibitemShut {NoStop}%
\bibitem [{\citenamefont {Peres}(2010)}]{Peres2010}%
  \BibitemOpen
  \bibfield  {author} {\bibinfo {author} {\bibfnamefont {N.~M.~R.}\
  \bibnamefont {Peres}},\ }\href {\doibase 10.1103/RevModPhys.82.2673}
  {\bibfield  {journal} {\bibinfo  {journal} {Reviews of Modern Physics}\
  }\textbf {\bibinfo {volume} {82}},\ \bibinfo {pages} {2673} (\bibinfo {year}
  {2010})},\ \Eprint {http://arxiv.org/abs/1007.2849} {arXiv:1007.2849}
  \BibitemShut {NoStop}%
\bibitem [{\citenamefont {Stauber}, \citenamefont {Peres},\ and\ \citenamefont
  {Guinea}(2007)}]{Stauber2007}%
  \BibitemOpen
  \bibfield  {author} {\bibinfo {author} {\bibfnamefont {T.}~\bibnamefont
  {Stauber}}, \bibinfo {author} {\bibfnamefont {N.~M.~R.}\ \bibnamefont
  {Peres}}, \ and\ \bibinfo {author} {\bibfnamefont {F.}~\bibnamefont
  {Guinea}},\ }\href {\doibase 10.1103/PhysRevB.76.205423} {\bibfield
  {journal} {\bibinfo  {journal} {Physical Review B}\ }\textbf {\bibinfo
  {volume} {76}},\ \bibinfo {pages} {1} (\bibinfo {year} {2007})},\ \Eprint
  {http://arxiv.org/abs/0707.3004} {arXiv:0707.3004} \BibitemShut {NoStop}%
\bibitem [{\citenamefont {Katsnelson}\ and\ \citenamefont
  {Novoselov}(2007)}]{Katsnelson2007}%
  \BibitemOpen
  \bibfield  {author} {\bibinfo {author} {\bibfnamefont {M.~I.}\ \bibnamefont
  {Katsnelson}}\ and\ \bibinfo {author} {\bibfnamefont {K.~S.}\ \bibnamefont
  {Novoselov}},\ }\href {\doibase 10.1016/j.ssc.2007.02.043} {\bibfield
  {journal} {\bibinfo  {journal} {Solid State Communications}\ }\textbf
  {\bibinfo {volume} {143}},\ \bibinfo {pages} {3} (\bibinfo {year} {2007})},\
  \Eprint {http://arxiv.org/abs/0703374v2} {arXiv:0703374v2 [arXiv:cond-mat]}
  \BibitemShut {NoStop}%
\bibitem [{\citenamefont {Wu}\ \emph {et~al.}(2017)\citenamefont {Wu},
  \citenamefont {Asselberghs}, \citenamefont {Contino}, \citenamefont {Chuang},
  \citenamefont {Politou}, \citenamefont {Radu}, \citenamefont {Huyghebaert},
  \citenamefont {Tokei}, \citenamefont {Soree},\ and\ \citenamefont
  {Heyns}}]{Wu2017a}%
  \BibitemOpen
  \bibfield  {author} {\bibinfo {author} {\bibfnamefont {X.}~\bibnamefont
  {Wu}}, \bibinfo {author} {\bibfnamefont {I.}~\bibnamefont {Asselberghs}},
  \bibinfo {author} {\bibfnamefont {A.}~\bibnamefont {Contino}}, \bibinfo
  {author} {\bibfnamefont {Y.-T.}\ \bibnamefont {Chuang}}, \bibinfo {author}
  {\bibfnamefont {M.}~\bibnamefont {Politou}}, \bibinfo {author} {\bibfnamefont
  {I.}~\bibnamefont {Radu}}, \bibinfo {author} {\bibfnamefont {C.}~\bibnamefont
  {Huyghebaert}}, \bibinfo {author} {\bibfnamefont {Z.}~\bibnamefont {Tokei}},
  \bibinfo {author} {\bibfnamefont {B.}~\bibnamefont {Soree}}, \ and\ \bibinfo
  {author} {\bibfnamefont {M.}~\bibnamefont {Heyns}},\ }\href@noop {}
  {\bibfield  {journal} {\bibinfo  {journal} {Graphene week poster
  presentation}\ } (\bibinfo {year} {2017})}\BibitemShut {NoStop}%
\end{thebibliography}%
\end{document}